# READINESS OF THE SOUTH AFRICAN AGRICULTURAL SECTOR TO IMPLEMENT IOT

In'aam Soeker, University of Cape Town, inaamsoeker@gmail.com

Shallen Lusinga, University of Cape Town, shallen.lusinga@uct.ac.za

Wallace Chigona, University of Cape Town, wallace.chigona@uct.ac.za

**Abstract:** As the world's population increases, so does the demand for food. This demand for food in turn puts pressure on agriculture in many countries. The impact of climate change on the environment has made it difficult to produce food that may be necessary to accommodate the growing population. Due to these concerns, the agriculture sector is forced to move towards more efficient and sustainable methods of farming to increase productivity. There is evidence that the use of technology in agriculture has the potential to improve food production and food sustainability; thereby addressing the concerns of food security. The Internet of Things (IoT) has been suggested as a potential tool for farmers to overcome the impact of climate change on food security. However, there is dearth of research on the readiness of implementing IoT in South Africa's agricultural sector. Therefore, this research aims to explore the readiness of the agricultural sector of South Africa for a wide implementation of IoT. This research conducts a desktop study through the lens of the PEST framework on the special case of South Africa. A thematic literature and documents review was deployed to examine the political, economic, societal and technological factors that may facilitate or impede the implementation of IoT in the agricultural sectors of South Africa. The findings suggest that the wide ranging political, economic, societal and technological constructs enable the implementation of IoT within South Africa's agricultural sector. The most important include current policies, technological infrastructure, access to internet, and mobile technology which places South Africa in a good position to implement IoT in agriculture.

## 1. INTRODUCTION

Southern Africa has been bedeviled by diminished and late rainfall and, long-term increases in temperatures. In 2015, South Africa has experienced the worst drought the region has had in decades. As a result, the food security of millions of people in the region has been jeopardized (NASA, 2019). Currently, the United Nations (UN) estimates that 690 million people suffer from hunger, with 250 million of them living in Africa, where malnutrition is growing fastest in the world (UN, 2020).

The accessibility and utilization of technology is known to enhance and facilitate the political, economic, social, and cultural growth of developing countries. The UN and World Bank have suggested that economic development is enhanced through investments in Information and Communication Technologies (ICT) (Ungerer, Bowmaker-Falconer, Oosthuizen, Phehane, & Strever, 2018; World Bank, 2018b). The role of technology has intensified in the lives of people as well as the countries ever since the COVID-19 pandemic hit (Ungerer et al., 2018; WSIS, 2020). COVID-19 has accelerated the capability of information sharing and in turn enhanced the efficiency of society (Rogers, Apeh, & Richardson, 2016; WSIS, 2020). Due to COVID-19 restrictions on movements, farmers who traditionally relied on face to face interactions with experts for their critical farming decision making were forced to retrieve information through digital channels such as phone calls, SMS, or voice messages (WSIS, 2020). Technology has shown its potential by enabling the availability and reliability of low-cost solutions that caters to all people, especially the vulnerable (Ungerer et al., 2018).

The Internet of Things (IoT) is an ecosystem that comprises of intelligent devices and machines, objects, people, or animals that are interconnected without human-to-computer or human-to-human interaction (Ayaz, Ammad-Uddin, Sharif, Mansour, & Aggoune, 2019). Previous studies have shown that IoT has emerged as an effective way for agriculture to improve operations, enable effective and efficient production and reduce the effect on the environment (Ungerer et al., 2018). Therefore, this raises the question what factors of readiness influences the agricultural sector to benefit from the implementation of IoT. Implementing IoT will, in the long-run, assist developing countries in Southern Africa to increase agricultural efficiency and encourages the sector to achieve sustainable developmental goals (Ungerer et al., 2018). However, there is paucity of research on the benefit of implementing IoT in South Africa's agricultural sector. As such, the research question for this paper is:

*"What factors of readiness influence the agricultural sector of South Africa to implement of IoT?"*

In answering this question, this study uses the Political, Economic, Social and Technology (PEST) analysis as a tool to evaluate the extent to which the agricultural sector is ready to implement IoT. The PEST analysis assesses IoT as a potential to increase the production and efficiency of the agricultural sector of South Africa and enable it to address its current climatic changes and food insecurity. The research conducts a desktop study utilizing a thematic literature and document review approach through the lens of the PEST framework.

South Africa is one of the leading countries in Africa in terms of technology use and adoption (AccessPartnership, 2018). The case of South Africa is representative of Southern Africa in terms of environmental conditions related to food production, while being an example of an aspiration to make use of IoT in improving the productivity of the agricultural sector in the Southern African region. IoT implementation within South African agriculture benefits the South African agricultural ecosystem, economy and society through improved agricultural production and efficiency, agricultural exports, and sustainability (Ayaz et al., 2019; Makate, Makate, Mango, & Siziba, 2019; Ungerer et al., 2018).

## 2. THE INTERNET OF THINGS

Kevin Ashton first coined the term Internet of Things in 1999, and defined it as a uniform way for the internet to connect and understand the physical world (Schoenberger, 2002). IoT is a network composed of physical integrated machines fitted with sensors, processors and networking systems that run either via the Internet or the on a local data network. Therefore, IoT can be described as the connection between the real world and the digital world. IoT enabled devices, with the assistance of sensors, collects data and appends it to the data of other smart devices connected to the network. The collected data can be analyzed and used to make decisions or archived for long-term purposes (De Cremer, Nguyen, & Simkin, 2017; Sullivan, 2018).

IoT generally refers to a mixture of technological functionality that intends to generate benefit based on functionality. For this reason, IoT devices range in complexity and are custom-made to achieve specific purposes. The flexibility of IoT ranges from a minimum requirement of possible communication and cooperation with other devices in the system to vastly complex. At a minimum communication and cooperation can occur via Bluetooth, Wi-Fi, or UMTS. To remotely control the smart devices, the devices are required to be addressable. The sensor technology enables the collection and transmission of information. While the processors and repository capacity allow the processing of collected data via the smart devices. The local usability of IoT extends to user interfaces such as smartphones (Makate et al., 2019). However, IoT is in its nascent stage and requires further improvement to become more accessible and less expensive to all. The benefits of IoT focuses on the ability of the network to gather and analyze data and, as a result, support organizations', communities, and individuals (Ayaz et al., 2019).

## 2.1. The Use of IoT in Agriculture

IoT provides an opportunity to transform various industries, including agriculture (Brewster, Roussaki, Kalatzis, Doolin, & Ellis, 2017). IoT technologies enables the agricultural sector to overcome food scarcity and climate concerns. This can be achieved through IoT devices that enable automated drip irrigation, weather forecasting, water level detectors, and soil level detection that use software intelligence, sensors and ubiquitous connectivity (Madushanki, Halgamuge, Wirasagoda, & Syed, 2019). These IoT devices link to sensors which produces and collects data that can be analysed using open software and, therefore, assisting farmers in managing and protecting crops by providing valuable information wirelessly or through low-powered networks (Brewster et al., 2017; World Food Programme, 2020). In addition, a benefit of IoT in agriculture is that it enables farmers to receive this information on their smart phones, which allows for efficient and effective agricultural farm management and increased productivity (Dlodlo, & Kalezhi, 2015).

## 2.2. The Use of IoT in South Africa

IoT was introduced into the South African economy more than a decade ago and continues to shape the country and influence most industries. Although it is a relatively new technology, South Africa has been using IoT for many years. Currently, there has been a nationwide production of network sensors used to connect everything including traffic controls and electricity grids as well as developing cloud infrastructure across the various sectors. In 2012, the South African National Roads Agency Limited introduced an e-tolling system, which charged vehicles for using the road; the system operated uses an IoT enabled systems. The IoT device senses e-tags or vehicle number plates as they pass. Compared to most of the African countries, South Africa has responded fast to IoT technologies. ESKOM, a South African public electricity utility company, has implemented smart meters that measure electricity usage and an IoT application that allows customers to identify the times, dates and areas that power-outages will affect (Onyalo, Kandie, & Njuki, 2015).

In South African agriculture, a grain producer in North West province, Jozeph du Plessis has been practicing precision farming on his dryland crops since 2001. du Plessis started precision farming using satellite imagery and yield monitors to rotate between soya beans and sunflowers. The precision software mapped and monitored the physical properties and chemical conditions of the soil. The spatial features was digitized, and computerized models measured the predicted yield of maize based on the soil-water holding capacity. Low potential areas were replaced with pastures and soil nutrient levels were analyzed and designed to reach optimal levels; therefore, the soil could be utilized to its full potential. In addition, du Plessis used a soil moisture meter to manage low water content, which determined the need for the implementation of a fallow system to allow cultivated areas of land to conserve water. The result of the technological innovation was that du Plessis was able to increase average maize yield and water usage (Ungerer et al., 2018).

The increased number of connected devices (IoT) and use of cloud-based services by the public, academia, business, and other sectors in South Africa has resulted in an increased demand for higher broadband and reliable connectivity (Department of Telecommunications and Postal Services, 2016). Sufficient internet bandwidth and economic infrastructure is essential for the implementation of IoT, therefore for Africa to implement IoT, governments are required to provide infrastructure that supports IoT solutions (365FarmNet, 2017; World Bank, 2018b; Zander, Trang, Mandrella, Marrone, & Kolbe, 2015). Furthermore, South African farmers continue to experience the effects of centuries of institutionalized racial discrimination as a result of Apartheid. The effect is mirrored in the distribution of agricultural land to racial minorities in commercial farming (Venter, Shackleton, Van Staden, Selomane, & Masterson, 2020; Cousins, 2016). Although, the benefits of IoT in agriculture are undeniable, it is not clear whether the agricultural sector in South Africa is ready to adopt and leverage the technology (Atayero, Oluwatobi, Alege, 2016).

## 3. ANALYTICAL FRAMEWORK

The PEST (Political, Economic, Sociological and Technological) framework is an analytical tool for examining how macro-environmental factors affect a phenomenon of interest. The framework is used to examine the success of a management initiative through understanding the relevant factors specific to its organizational environment. The organizational environment consists of the external social and physical factors that influence its decision-making process (Duncan, 1972). PEST assumes that these external factors, as well as indirect conditions, shape the organizational environment by influencing its value-adding capabilities. Therefore, an aerial perspective of the external environment is produced through a PEST analysis. This becomes increasingly important when trying to refine larger environments to understand its organizational systems (Law, 2006).

The PEST framework was used in this study to produce an understanding of the context through analyzing the PEST constructs of the agricultural industry within its natural environment. In addition, the framework helped us to narrow the focus of the study to specific factors of the PEST framework that influence readiness. This, in turn, developed meaningful insights and in-depth understandings of the sectors readiness for IoT implementation. However, each dimension of the PEST framework creates a multitude of variables. Therefore, the framework is viewed through an information systems lens to determine which variables have the greatest impact on the agricultural sector.

PEST seeks to assess the political, economic, social, and technological conditions of the agricultural sector to understand the factors of readiness of the industry to implement IoT. PEST addresses the most common issues of IoT in the agricultural sector, to determine whether there is benefit in implementing IoT in the South African agricultural sector.

By assessing the sector through the four constructs of the framework, the current and prospective state of readiness in the agricultural sector is understood. PEST analysis is used to analyze the macro-factors (political, economic, sociological, and technological) of the external environment to enable strategic management through leveraging its opportunities and mitigating threats (Law, 2006; Duncan, 1972). The results of the PEST analysis will illustrate the environment within the agricultural sector as well as identify the aspects that may strengthen, impede or create opportunities for technological growth.

## 4. RESEARCH METHODOLOGY

The study employs a desktop study using a thematic literature and documents review based on the PEST framework. The exploratory nature of this study enabled a holistic understanding of the research question. The study used a number of bibliographic references, which the researcher checked and compared. These sources were critiqued and combined to develop standpoints and arguments for the PEST framework. In addition, statistical data and grey literature were used to determine the background of the country.

The research was conducted through an analysis of the contributions made of the political, economic, socio-economic, and technological purposes of IoT within the agricultural sector of South Africa over the period of 2015 – 2020. The period is representative of the most recent uses of IoT within South Africa. The documents were retrieved between January and September 2020. We searched for the document using the following key words: Internet of Things, IoT, agriculture, IoT agriculture, IoT agriculture South Africa, South Africa, PEST analysis, South Africa PEST analysis. The key words were chosen due to its representativeness of the main objective of the research. The contributions were searched in two search engines, Google and Google Scholar, and multiple digital libraries to collect information. The chosen digital libraries were Elsevier, Xplore Springer, IEEE, and IGI which were chosen based on their contributions.

Initially, the search yielded 150 peer-reviewed papers. The papers were screened on the basis of titles, keywords, exclusions, and full articles which reduced it to 75. In the end, 26 papers out of the 150 were used. In addition, South African policy documents and international statistical documents were used. All research contributions found on the theme of the PEST factors of IoT in South African agriculture were gathered, assessed, and analyzed to understand the research topic.

## 5. FINDINGS AND DISCUSSION

### 5.1. Case Context

South Africa is a rich and diverse county with an agricultural sector that is characterized by its dual agricultural economy (AgriSETA, 2019; Mordor Intelligence, 2017; Goldblatt, 2016). The diversity of the South African agricultural sector is depicted by various types of farming, such as, crop and animal production, horticulture, and dairy, fish and, game farming (Mordor Intelligence, 2017). The agricultural economy of South Africa is categorized into well-developed commercial farming, which makes up approximately 87% of agricultural land, and subsistence, or small-scale farming, which occupies only 13% of total agricultural land (Tibesigwa, Visser, & Turpie, 2017). However, there is a lot of complexity and fluidity between commercial and subsistence farming (AgriSETA, 2019; Goldblatt, 2016). South African commercial farming is highly developed, and export driven and is still a significant exporter of agricultural products. Whereas small-scale subsistence farming is less developed and subject to various resource and production constraints (Kushke, I., & Jordaan, 2017). As with many sectors in South Africa, the agriculture sector is affected by race. Where most agricultural employees in the agricultural sector are predominantly black (65%), followed by colored (19%), white (12%) and very few Asians (1%) (AgriSETA, 2019; Statistics South Africa, 2020).

Only 13.7% of South Africa's total surface area is arable and used for crop production, while only 3% of the arable land is considered as truly fertile land (Goldblatt, 2016; Jacobs, Van Tol, & Du Preez, 2018). Since the early 1990's, South Africa was left with less than two-thirds of the number of farms it had due to climate change, water scarcity and declining farming profitability (Goldblatt, 2016; Jacobs et al., 2018). The lost farms have been used for other uses, while the remaining farms have increased its productivity to achieve sustainability through technology adoption within their farming practices (Kane-Berman, 2016).

### 5.2. Political Factors

South Africa has experienced immense economic and social reform since the transition into a democratic state in 1994. Following the end of the Apartheid-era, economic reform and policy changes proposed to eliminate the previous Nationalist Governments' socialist control of agriculture by improving the agricultural environment and redressing land inequalities (Goldblatt, 2016; Ungerer et al., 2018). The government's approach to achieving social, political and economic reform and integration has undoubtedly shifted, and ICT policy needs to adapt and respond accordingly. Since 1990, focus was placed on redressing the injustices of apartheid and decades of racial discrimination through the Reconstruction and Development Programme (1994) (Goldblatt, 2016).

New approaches to ensure equality, quality of life and poverty eradication are targeted through ICT policy (National Treasury, 2019). The aim for equality within the agricultural sector arises from South Africa's political history. Most commercial farmers were found to be white with access to resources, which enables the implementation of IoT. While most subsistence farmers were found to be black with a lack of access to resources which makes them less likely to implement IoT (AgriSETA, 2019; Statistics South Africa, 2020). This leads to an uneven distribution of IoT readiness in the agricultural sector which hinders the aim of ICT policy to ensure equality as readiness is not uniform across the sector and remains divided on race. Therefore, IoT can be implemented; however, its implementation will not align and redress currently inequalities in the country (KMPG, 2012; UN, 2020).

Current policies show the need for IoT in the South African agricultural sector. The National Development Plan 2030 policy highlights the importance of initiatives that links the agricultural sector to the green economy. The Department of Agriculture, Forestry and Fisheries Integrated Growth and Development Plan 2012 emphasizes sustainable agriculture with an aim to benefit all South Africans. These policies create the demand for IoT in terms of enabling a greener and more sustainable agricultural sector, with less water, fertilizers, and pesticides usage. The political aim for South African agriculture is conducive to the implementation of IoT because IoT enables sustainability, productivity and, information sharing within the sector (Ungerer et al., 2018). However, there is not much evidence on the effect and progress of these policies.

The lack of farmer support is causing the number of farms to decrease. The African Farmer's Association of South Africa (AFASA) reported that of its total members, only one third of them farm for the income, of which only 2%, of those farming commercially, are successful (Nayak, Kavitha, & Rao, 2020). Therefore, regardless of the type of farming, many farms in South Africa are operating hand-to-mouth and require support to become sustainable. Over the years, the number of commercial farms has substantially decreased. However, this decrease in numbers has been accompanied by increases in farm sizes as well as the implementation of various technologies on the farms, including IoT (AgriSETA, 2019; Ungerer et al., 2018). Government policies that support farmers financially creates a demand for IoT by providing commercial and subsistence farmers equal opportunities for technological innovation. Therefore, subsidies given to farmers enables farmers readiness to implement IoT and provides an opportunity for actual IoT implementation on their farms (Ayaz et al., 2019; Ungerer et al., 2018).

### 5.3. Economic Factors

The South African forecasted gross domestic product (GDP) showed a reduction of 2% in GDP from 2019 – 2020 due to the COVID-19 outbreak, causing additional economic challenges to SA's increasing debts, large fiscal deficits, depressed growth, and high social vulnerabilities (Statistics South Africa, 2020). While the agricultural sector has had a good quarter with a rise of 27,8% in production activity due to an increase in the production of field crops, animal and horticultural products (Statistics South Africa, 2020). This economic growth calls for the development of technological strategies that further improves the sector (Ungerer et al., 2018; World Food Programme, 2020). Therefore, the economic state of the agricultural sector can be seen to be in a good position to implement IoT strategies to further increase and improve productivity in the sector (Ayaz et al., 2019; AgriSETA, 2019; Ungerer et al., 2018; World Food Programme, 2020).

South African economic growth has been accelerated by the agricultural sector. To further improve such growth, the agricultural focus should shift to becoming more competitive through technological innovation. To become more competitive, farmers would need to implement technologies that improves productivity. IoT provides the platform for more competition within the sector, through improving productivity and, thereby, improving agriculture exports (Ayaz et al., 2019; Ungerer et al., 2018).

The overall slow growth rate and high inequality levels strengthen each other. The inequality promotes resource contestation (through corruption and crime), which impedes investment required to facilitate equality and technological innovation (Western Cape Government, 2017). Farmers who implement IoT are at risk of criminals stealing IoT devices in open fields (Ayaz et al., 2019). In addition, the corrupt South African government that might steal funding intended for agricultural technological development hinders farmers. This affects the implementation of IoT by farmers, as they are not provided with the financial support or security from government, especially those farmers who are in weaker positions, such as those doing subsistence farming (World Bank, 2018a; World Bank, 2018b).

South Africa has been known for its slow uptake and access to ICT compared to other countries (Onyalo et al., 2015), while most of its progress has been due to increased competition from challenger firms (CCRED, 2016). Technology is being used to facilitate successful land reform as illustrated by modern crowdfunding platforms, while increasing agricultural economic development opportunities in rural areas (World Food Programme, 2020. The ICT policy enables equal opportunity for IoT implementation across both urban and rural areas, thereby, offering opportunity for growth within the sector (AgriSETA, 2019; Goldblatt, 2016)

## 5.4. Social Factors

Overall, the worldwide trend is unemployment is increasing with the implementation of technology leading to a reduction in the demand for semi-skilled and unskilled labor (World Food Programme, 2020). The lack of economic growth in South Africa negatively impacts the employment rate in the agricultural sector; as a result, there was a 0.6% decrease in employment due to the increased mechanization of farming processes (Statistics South Africa, 2020). The high unemployment rate may cause government resistance for IoT initiatives; however, there is no evidence to support this. The rapid technological changes in agriculture production is affecting skill-intensive employment demands through the creation of biases. The discrimination against the South African labor force is leading to increased unemployment. However, the agricultural industry is seen as a passage to growth in the labor and sociological sector. Technological innovation within the agricultural sector has promoted agrarian transformation and improved agricultural production and, therefore, should be supported (National Treasury, 2019).

A major improvement in South Africa's sociological capacity would entail breaking free from high unemployment that the region has been stuck for decades. IoT is considered to both increase and decrease employment rates in agriculture. The former is achieved through attracting the youth to the sector through the use of technology, in the fourth industrial revolution. The latter is a result of IoT devices replacing workers who do on-field manual labor tasks that can be performed through IoT implementation (Ayaz et al., 2019). Skilled laborers able to use new technologies will need to be accommodated with equitable wages to compensate for the need for a new skill set. While further accommodation is needed to upskill, retrain and socially develop current unskilled workers technological competence (World Food Programme, 2020). The reduction in labor costs associated with IoT acts an incentive for farmers to implement IoT in their current agricultural practices (KMPG, 2012). However, the costs associated with upskilling workers may act as a hinderance to IoT implementation.

Education levels of farmers play a key role in the implementation of IoT in the more educated farmers may be more ready to implement IoT. Whereas, those who are less educated may be more resistant to implement an unknown technology. In South Africa, commercial farmers are predominantly white and, therefore, more educated and likely to implement IoT. While subsistence farmers are mostly black, and more likely to be uneducated and less inclined to implement IoT. This leads to inequality across the sector, even though farmers are ready to implement IoT (Ayaz et al., 2019; Brewster et al., 2017; Jacobs et al., 2018).

The age of employees in the agricultural sector was reported to be between 15 and 65 year, while 40% were under the age of 35, 52% were between the ages of 35 to 55 and 8% were found to be older than 55 (AgriSETA, 2019). The large proportion of young farmer employees may act as an incentive for farmers to implement IoT in their practices. However, the education level of the young farmers may act as a hinderance to the use of IoT. With a larger distribution of the youth using mobile technologies, the transition to IoT will be easier if farming practices are managed via the mobile phone (ICASA, 2020).

Further consideration regarding the key sociological factors that affect the agricultural sector in South Africa are ethics and privacy issues associated with IoT. Farmers may be resistant to storing their agricultural information on IoT devices that may be vulnerable to security breaches by competitors. This may cause farmers to perceive IoT as unattractive (AgriSETA, 2019; Ungerer et al., 2018; World Food Programme, 2020). If these issues are left unresolved, they will contribute to a decreased uptake of IoT on farms and an increased migration of farmers to urban areas as a result of unprofitable farming.

### 5.5. Technological Factors

Agriculture in South Africa is dependent on farming equipment, labor and infrastructure. The challenge would be optimizing the accessibility to resources such as fertilizers, seeds, and suitable cultivars, parallel to management techniques and expertise of agricultural processes and applications. This can be addressed through the use of IoT because it enables the optimal use of these farming inputs by farmers. In addition, it provides farmers with farming management capabilities that allows farmers to make informed decisions regarding crop production, thereby increasing production and profitability (Ayaz et al., 2019; Ungerer et al., 2018). The issues faced in South Africa are the accessibility to IoT, the organization of internet access, and the integration of IoT sensor technologies (Nayak et al., 2020). Access to new technologies remains unevenly distributed across commercial and subsistence agriculture in South Africa. Lack of access to technology is one of the major inhibitors for the implementation of IoT by subsistence farmer, thereby, increasing technological inequality in the sector (Ayaz et al., 2019; Jacobs et al., 2018).

Most farming equipment in operation are analogue and not compatible with new technologies and networks. However, modern agricultural equipment enables data collection and analysis by allowing farmers to meet requirements and improve their agricultural processes. Technological solutions already exist to solve these issues. For instance, GPS systems and Bluetooth together with software, interoperability and, standardization ensure that legacy machinery can be digitalized (World Bank, 2018b). The right IoT technology can be seamlessly integrated into existing farming equipment and linked to mobile technology. This may lead to greater implementation of IoT in agriculture. Commercial farmers are opting for more high-tech equipment, putting them in a better position to implement IoT technologies in their farming practices. While subsistence farmers lack access to advanced farming technologies, therefore, reducing their ability to use IoT (KMPG, 2012).

Technological infrastructure continues to be inadequate in rural areas. For certain technological solutions to operate, an internet connection is necessary. Like other sectors of the economy, modern agriculture relies heavily on effective telecommunication infrastructure. South Africa's internet penetration stood at 62% in January 2020 (ICASA, 2020). In rural areas of developing countries, limited telecom infrastructure is common which causes challenges for the full utilization of IoT and realization of its cost-effectiveness (365FarmNet, 2017). However, in South Africa the lowest rural internet coverage was found in the Northern Cape with a coverage of 3G and LTE sitting at 99%, 97% and 72% in 2019, respectively. While the highest coverage was found in the Gauteng Province sitting at 100%, 100% and 99%. Urban internet coverage is marginally better than rural internet coverage. In 2019, all nine provinces had 100% 2G urban coverage, the Northern Cape Province had the lowest 3G urban population coverage at 99% in 2019, with the rest of the other Provinces had 100%. The Northern Cape had the lowest LTE urban coverage at 98% in 2019 (ICASA, 2020). This shows that South Africa's technological infrastructure in the agricultural sector is ready for IoT implementation.

With an increase use of technology, so comes the rise in demand for data and reliable broadband technology. In January 2020, two undersea cables broke which resulted in slow internet connection across the country. This caused an inconvenience to many South Africans who were working from home as a result of the COVID-19 pandemic (Business Insider South Africa, 2020). Although these issues are uncontrollable, South Africa's internet infrastructure has improved and, therefore, places them in a good position to support IoT implementation throughout the agricultural sector.

The Department of Telecommunications and Postal Services (DTPS) is in the process of deploying telecom infrastructure to enable more accessible fixed-line broadband access as it provides access to unlimited data transmission (National Treasury, 2019). Furthermore, it has been claimed that mobile broadband is not a sustainable technology for the future of broadband due to its limited data transmission (National Treasury, 2019). However, in developing countries, where mobile technology and Wi-Fi are highest in demand in the agricultural sector, remains one of the main tools used to penetrate technological innovation (Gokul, & Tadepalli, 2017). In South Africa, smartphone penetration increased from 81.7% in 2018 to 91.2% in 2019 (ICASA, 2020). The widespread use of mobile technology created an opportunity for IoT use within the agricultural sector.

Mobile technologies are used as a tool in IoT to monitor water, crops, and land resources. Third generation (3G) connectivity is an accessible option for farmers because it is available across most of South Africa (Ayaz et al., 2019; World Food Programme, 2020). In South Africa, mobile network coverage is in a good state with national population coverage for 3G increasing from 99.5% in 2018 to 99.7% in 2019 and national population coverage for 4G/LTE increasing from 85.7% in 2018 to 92.8% in 2019 (ICASA, 2020). Data collected from monitoring crops through IoT devices can be transmitted via SMS's, over the slowest networks, therefore, existing ICT infrastructure can be used. However, due to high mobile data costs and limited internet speed, low-powered, short-range networks or low-rate wireless PAN (LoRaWAN) can be used as an alternative to Wi-Fi and 3G, especially for the utilization of IoT (Madushanki et al., 2019). Therefore, the wide range of broadband options reduces the technological inequality within the sector because it enables the use of IoT across different network capabilities and provides efficient connectivity over a low-cost and reliable spectrum (Nayak et al., 2020).

# 6. CONCLUSION

Food security and sustainable food production has been long been recognized as a global issue. Evidence exists that the use of technology in agriculture has the potential to improve food production and sustainability. Therefore, the aim of this study was to explore the readiness of the agricultural sector of South Africa for a wide implementation of IoT. Through the PEST analysis, this study developed an in-depth understanding of South Africa's agricultural context in terms of political, economic, social, and technological factors. The use of the PEST framework in this research enabled the assessment of South Africa's readiness to implement IoT. The study found that the introduction of IoT within South African agriculture has many challenges for a wide implementation, however, there exists some degree of readiness. Overall, this study observed that some factors place South Africa in good position to implement IoT, such as, increased farmer support from South African government, the willingness of younger farmers to implement IoT, reduction in labor costs, easy integration of IoT technology to mobile phones, and wide ranging internet coverage. While, some challenges were found to be an unequal distribution of IoT technology between black and white farmers, lack of access to technology, risk of criminal activity, influence of age and education on the willingness to adopt IoT, slow uptake and access to ICT, and the potential to increase unemployment rates. However, current policies show that the South African government is actively planning to overcome these issues. There were some limitations to this study and so interpretation and generalization must be undertaken cautiously. The exploratory nature of this study requires that more cases are included in this investigation. In addition, this study would have benefited greatly from the inclusion of both qualitative and quantitative empirical data collected from the relevant parties in the agricultural sectors of Southern African countries. An extension to this work could consider assessing the perceptions of farmers to implement IoT in agricultural sector of South African agricultural.